# Crystal Structure of Gold Hydride

Valentina F Degtyareva

Institute of Solid State Physics, Russian Academy of Sciences,
Chernogolovka 142432, Russia

E-mail: degtyar@issp.ac.ru

**Abstract.** A number of transition metal hydrides with close-packed metal sublattices of fcc or hcp structures with hydrogen in octahedral interstitial positions were obtained by the high-pressure-hydrogen technique described by Ponyatovskii et al, Sov. Phys. Usp. 25 (1982) 596. In this paper we consider volume increase of metals by hydrogenation and possible crystal structure of gold hydride in relation with the structure of mercury, the nearest neighbor of Au in the Periodic table. Suggested structure of AuH has a basic tetragonal body-centered cell that is very similar to the mercury structure Hg - $tI2$. The reasons of stability for this structure are discussed within the model of Fermi sphere – Brillouin zone interactions.

*Keywords*: Metal hydrides; Volume changes; Crystal structure

## 1. Introduction

Transition metals usually form hydrides with close-packed atomic arrangements (face centered cubic, *fcc*, or close-packed hexagonal, *hcp*) of metal atoms and interstitial arrangements (tetrahedral and/or octahedral) of hydrogen in the metallic host lattice. The experimental method of high-pressure high-temperature synthesis of metal hydrides is described in [1,2]. With this method, several hydrides MeH (Me - transition metals and transition metal alloys) have been synthesised in a gaseous hydrogen atmosphere at high pressure and temperature. Structural investigations of MeH revealed an increase of volume related to a metallic atom.

An interesting observation is that this volume increase is approximately the same for different metals ranging from ~2 to ~3 Å$^3$ per metal atom for the composition MeH. The volume expansion in metal hydrides was a subject of many discussions [2-6]. At first glance one could expect that the hydrogen-induced expansion of the metal lattice should be related to the ratio of the metal and hydrogen radii, the latter being assumed equal to ~0.6 Å. However, experiment shows that the volume expansion is independent of the radii ratio and is larger than expected from the simple model of hard sphere packing assuming the occupation by small spheres (H) of the interstitial positions formed by large spheres (Me). Experimentally observed $\Delta V$ (Å$^3$) in transition metal hydrides [2] are listed in the Table 1, where $\Delta V$ refers to one atom H. Note that the structures of all hydrides are *hcp* or *fcc*, except FeH with a double hexagonal close-packed structure (*dhcp*) [7]. In this paper, the volume increase due to hydrogenation is discussed in connection with the changes in the number of valence electrons accompanying the formation of hydrides.

An attempt to synthesis gold hydride was reported in [8] by annealing of pure gold in a hydrogen atmosphere at ~50 kbar and ~400° C. The composition of synthesized material was estimated as AuH and an orthorhombic lattice was suggested. In this paper, we consider possible crystal structure of AuH in relation with the structure of mercury, the nearest neighbor of Au in the Periodic table. This suggestion is supposed assuming the atomic volume increase due to hydrogenation is nearly equal to the volume increase by moving elements along the row in the Periodic table. The reasons of stability for this kind of structure are discussed with consideration Fermi sphere – Brillouin zone interactions within the nearly-free electron model.

## 2. Results and discussion

*2.1. Correlations between the volume and hydrogen content of metal hydrides*

The observed volume increase ΔV in metal hydrides can be understood by consideration how the electron of hydrogen is involved in the valence band of the host metal consisted of *d* and *s* parts. Magnetic measurements on hydrides of transition metals have shown an increase of *d* -band filling by ~0.5 electron per H atom [1-3]. The *s*-band filling is also expected to increase. The increase in the number of *s* valence electrons is responsible for the increase in the atomic volume of metal hydrides. Atomic volumes of transition metals are nearly insensitive to the filling of the *d* -band and the volume increase correspond to addition of ~0.5 *s*-electron per H atom.

The value of volume expansion per one *s* valence electron can be estimated from the volume relations in the post-transition (or *sp*) elements shown in Fig. 1 [9]. These elements have filled *d*-shells, which form ionic cores with the nearly equal size for the elements from one row of the Periodic table. For the elements in one row, the atomic volumes increase nearly linearly with increasing group numbers equal to the numbers of valence *sp* electrons. The change in the atomic volume is associated with the number of *sp* outer electrons, as was considered by Schubert and called the "valence electron volume" [10].

Estimations from plots on Fig. 1 give the volume per one *s* electron ~4 to ~5 Å$^3$. Thus, the volume increase ΔV ~ 2 Å$^3$ in transition metal hydrides corresponds to ~0.5 electron increase in the *s*-band as expected from experimental studies [1-3].

It is interesting to compare volumes of PdH and the neighbouring to Pd element Ag. Both volumes are equal to ~17.1 Å$^3$ that points to the nearly equal number of valence *s* electrons in PdH and Ag. Recently, a platinum hydride has been synthesized under high hydrogen pressure. The hydride had a close-packed hexagonal structure [11,12] and lattice parameters $a = 2.779$ Å and $c = 4.731$ Å at $P = 30.5$ GPa. The volume difference between the platinum hydride and Pt metal was ΔV= 2.09 Å$^3$ and therefore corresponded to the PtH composition [12]. It should be noted that PtH is structurally similar to the neighboring to Pt element – gold, which accepted the *hcp* structure at pressures above 250 GPa [13].

A special case is the AuH hydride reported to be synthesised by Antonov et al. [8]. Elements Cu, Ag and Au continue transition metal rows and have their *d*-band already filled. Therefore the additional electron from hydrogen dissolved in these metals should go to the *s*-band. One could expect that the addition of one valence electron to Au will make the atomic volume and crystal structure of AuH similar to those of Hg, the neighbouring element of Au in the Periodic table.

*2.2. Suggested structure for gold hydride*

The position of gold in the Periodic table is in the group IB right after the transition metals. The electron energy levels of gold correspond to the completely filled 5$d^{10}$ band and one electron is on the 6*s* level. The crystal structure of gold is face-centered cubic, the same as that of the neighbouring transition metals (Ir, Pt) and group IB metals (Cu, Ag). The elements of the IIB group have considerably different structures. Particularly, Hg has a rhombohedrally distorted *fcc* structure at normal pressure and a tetragonally distorted *bcc* structure at pressure above 3.4 GPa. The latter structure of Hg (space group *I4/mmm*, two atoms in the unit cell, Pearson symbol *tI*2) can be retained at atmospheric pressure and at 77 K and has $a = 3.995$ Å; $c = 2.825$ Å,; $c/a = 0.707$; atomic volume 22.54 Å$^3$ [9].

An analysis of the diffraction pattern for AuH (Fig. 3) revealed the possibility of indexing the group of strong diffraction peaks on the base of a tetragonal body-centered cell with lattice parameters $a = 3.933$ Å and $c = 2.782$ Å, $c/a = 0.707$, atomic volume 21.5 Å$^3$ assuming two atoms in the unit cell. This structure is very similar to the mercury structure Hg-$tI$2 as compared in Table 2. The pure gold has the atomic volume 17.0 Å$^3$, so the volume increase due to the formation of hydride is 4.5 Å$^3$, which is approximately twice that for the transition metal hydrides. This agrees with the assumption discussed above that the electrons supplied by the hydrogen atoms all go to the *s*-band of the metals like Au with the already filled *d*-band, but are shared in approximately equal quantities of ~0.5 el. per H atom between the *d*- and *s*-bands in the transition metals, in which the *d*-band is partly empty.

In the diffraction pattern of AuH, there are a few unindexed weak peaks indicating that the real structure of AuH is more complex than proposed above. This structure is likely to be represented by a supercell on the base of the *tI*2 lattice, in which the Au atoms are slightly shifted from the ideal positions due to the ordered interstitial occupation by the hydrogen atoms. The structure of AuH therefore needs further refinement. Nevertheless, the suggested basic structure for AuH gives some confirmation that hydrogen content of the gold hydride is close to MeH.

It seems also worth noting that the formation of the gold hydride is accompanied by the change in the color of the sample from gold-yellow to silver-gray indicating a substantial change in the valence electron state of Au caused by the hydride formation. The hydrogenated sample is growing up in volume with some cracks. These changes correspond to volume increase by ~25% as should be expected by formation AuH. Within several days at ambient conditions the color of the sample returned gradually from gray to gold-yellow with volume shrinking.

A question arises what determines the stability of the tetragonally distorted body-centered structure of Hg and AuH. The model of Fermi sphere - Brillouin zone interactions points to the important role of the electronic contributions in the structural stability of metals with the *s* or *sp*-type valence electrons [14]. Fig. 3 shows the Brillouin zone for the mercury structure Hg - *tI*2, constructed from the first Brillouin planes of the (110) and (101) types and the corresponding Fermi sphere for z = 2 of the valence electrons. The zone was constructed using the specially developed program BRIZ [15]. The Hg-*tI*2 structure is stable due to the closeness of the Fermi surface and the (101) planes of the Brillouin zone characterized by the wave vector q.

Formation of the energy gap at the condition $k_F \approx$ ½ q results in the increase of the electron density of states, which leads to the decrease of the kinetic energy of the electrons and therefore to the decrease of the total energy of the crystal. The tetragonal distortion of the *bcc* structure brings the Brillouin planes (101) into contact with the Fermi sphere and thus to the decrease in the electron energy. This model explains the stability of the gold hydride under the assumption that its number of valence electron is z = 2. The Fermi sphere – Brillouin zone interactions were earlier shown to determine the formation of distortions and superlattices observed in several phases based on *bcc* and *hcp* structures [16,17]. This model allows for estimations of valence electron counts to satisfy the Hume-Rothery rules for phase stability in alkali and alkali-earth metals under compression [18,19].

Recent attempts to prepare new samples of gold hydride were failed and we had to use the X-ray diffraction patterns measured long ago. Recent high-pressure studies of interactions of noble metals (Cu, Ag and Au) with molecular hydrogen [20] did not reveal the formation of any gold and silver hydrides at pressures up to 113 and 87 GPa, respectively. In these experiments was produced a new *hcp* copper hydride at 18.6 GPa and confirmed the formation of the *fcc* copper hydride at 12.5 GPa in accordance with earlier results [21]. The hydrogen content of the copper hydrides was estimated as H/Cu = 0.4 [21] and H/Cu = 0.5 [20] from the hydrogen-induced increase in the unit cell volume of the metal. The estimate was based under the assumption that the hydrogen dissolved in copper expands its lattice by ~2 Å$^3$ per H, as in the case by hydrogenation of transition metals. According to the discussion above, the volume effect of hydrogenation of the noble metals should be approximately twice as large compared to the transition metals, therefore a more feasible estimate of the hydrogen content of the copper hydrides is H/Cu ~ 0.2–0.3. Relatively high increase in volume expected by formation of noble metal hydrides assumes not to go high with pressure but rather it is necessary the subtle combination of T-P conditions. That is a reason why Au hydrogenation was not observed in experiments with the Au-H mixtures by increasing pressure up to ~100 GPa [20].

## 3. Summary

The long-standing problem of volume expansion in the metal hydrides can further be understood under the assumption that it is mainly governed by filling the *s*-band of the metal with electrons supplied by the hydrogen atoms. The "valence electron volume" can be estimated from the dependences of the volume of the post-transition *sp* elements vs. their atomic number (see Fig. 1). The "valence electron volume" is approximately twice as large as the volume increase per one hydrogen atom dissolved in a transition metal. This result well agrees with the earlier findings [1–3] that approximately one half of the electron from each H atom absorbed by the transition metal fills up its *d*-band and another one half goes to the *s*-band.

The proposed crystal structure of the gold hydride, reported producing in a hydrogen atmosphere at high-pressure high-temperature conditions, is considered as closely related to the Hg-*tI*2 structure of mercury, the nearest gold neighbor in the Periodic table. The transformation of the *fcc* crystal structure of gold to the structure of mercury implies that the absorbed hydrogen increased the number of valence electrons in the Au metal to $z = 2$. This is an example of the modern "alchemistry".

**Acknowledgments**

The author gratefully acknowledges Dr. V.E. Antonov for valuable discussions. This work was supported by the Program "The Matter under High Energy Density" of the Russian Academy of Sciences.

**Table 1**. Volume increase accompanying the formation of monohydrides of metals of the 3$d$ and 4$d$ rows (results from Ref. [2])

| MeH | cell | $\Delta V, Å^3$ | MeH | cell | $\Delta V, Å^3$ |
|---|---|---|---|---|---|
| CrH | hcp | 2.2 | MoH | hcp | 2.2 |
| MnH | hcp | 1.8 | TcH | hcp | 2.0 |
| FeH | dhcp | 1.9 | - | - | - |
| CoH | fcc | 1.8 | RhH | fcc | 2.4 |
| NiH | fcc | 2.2 | PdH | fcc | 2.4 |

**Table 2**. Comparison of the crystal structure Hg-II (space group $I4/mmm$, two atoms in the unit cell, Pearson symbol $tI2$ [9]) and the suggested subcell of AuH

|  | $a$, Å | $c$, Å | $c/a$ | $V_{at}$, Å$^3$ |
|---|---|---|---|---|
| Hg-II at 77K | 3.995 | 2.825 | 0.707 | 22.54 |
| AuH (subcell) | 3.933 | 2.782 | 0.707 | 21.5 |

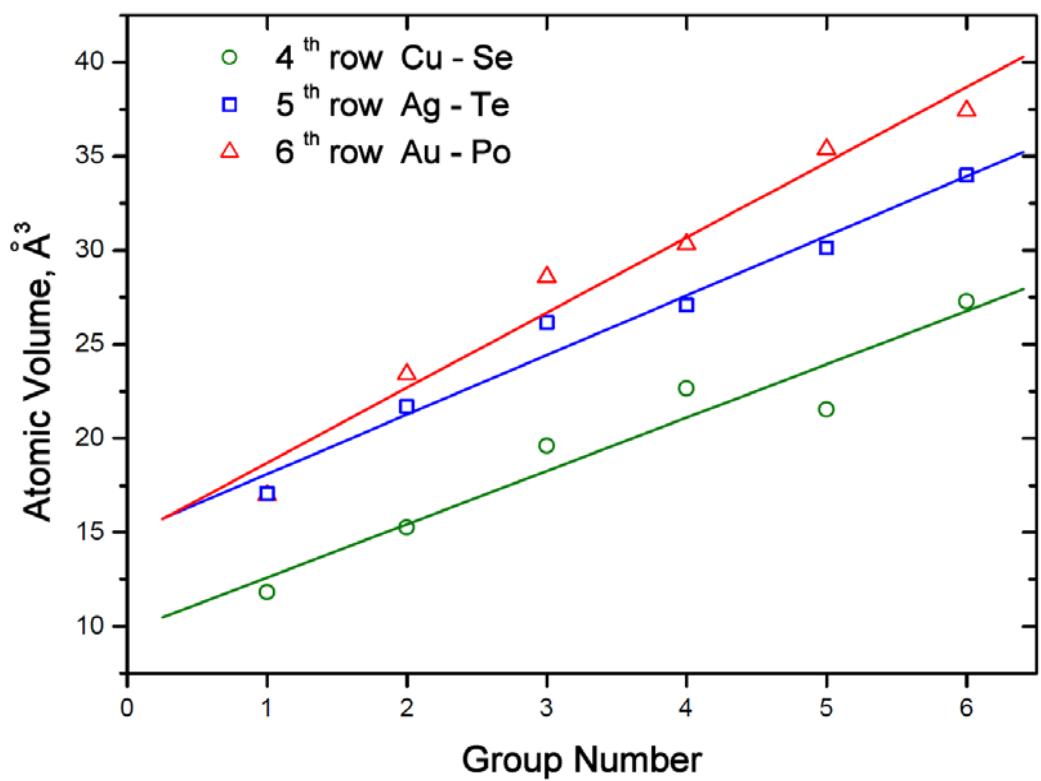

Fig. 1. Atomic volumes of the *sp* elements of 4-th, 5-th and 6-th rows of the Periodic table; structural data are taken from [9]. Nearly linear increase of the atomic volume along one row demonstrates the nearly constant "valence electron volume" (the volume per one valence electron).

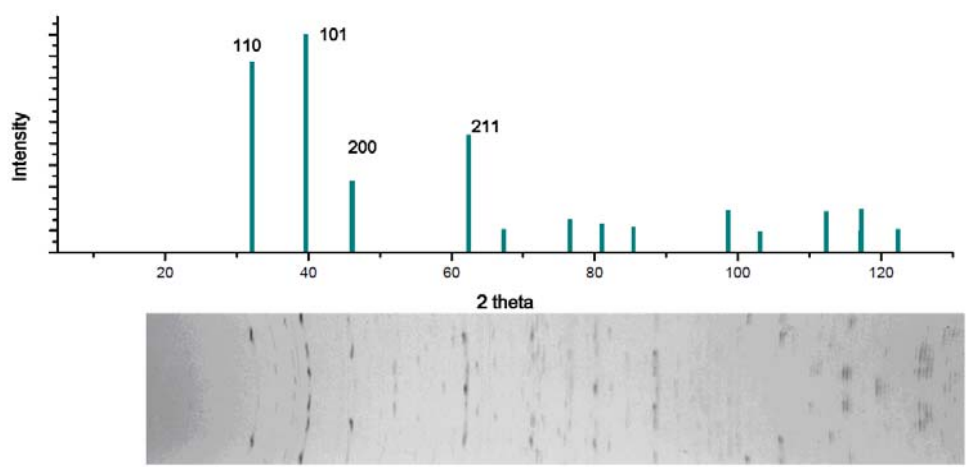

Fig. 2. Experimental diffraction pattern of one of the Au sample hydrogenated under high pressure (bottom of the figure) and a calculated pattern for the tetragonal structure *tI*2 similar to that of β-Hg with the lattice parameters given in Table 2. Cu-Kα radiation, room temperature.

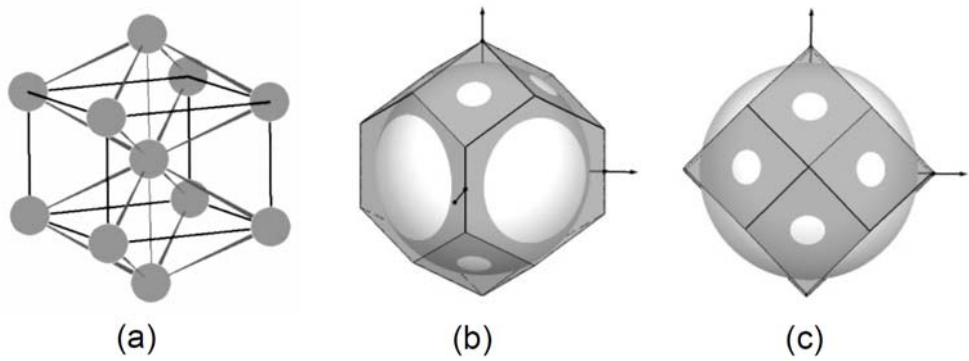

Fig. 3. (a) The crystal structure of Hg-*tI*2; the coordination number for a metal atom is equal 10. This structure is considered as the basic cell for AuH. (b,c) The Brillouin zone for the structure Hg-*tI*2 with the planes of the (110) and (101) types: the common view (b) and the projection along [001] (c). The inscribed sphere is shown with the Fermi radius $k_F = (3\pi^2 z/V)^{1/3}$ for the number $z = 2$ of valence electrons.